\documentclass[a4paper]{jpconf}
\usepackage{graphicx}
\usepackage{color}
\newcommand{\mincir}{\raise
-3.truept\hbox{\rlap{\hbox{$\sim$}}\raise4.truept\hbox{$<$}\ }}
\newcommand{\magcir}{\raise
-3.truept\hbox{\rlap{\hbox{$\sim$}}\raise4.truept\hbox{$>$}\ }}
\newcommand{\minmag}{\raise
-3.truept\hbox{\rlap{\hbox{$<$}}\raise5.truept\hbox{$<$}\ }}
\newcommand{\be}{\begin{equation}}
\newcommand{\ee}{\end{equation}}
\newcommand{\ba}{\begin{eqnarray}}
\newcommand{\ea}{\end{eqnarray}}

\begin{document}

\title{Alternative High-$z$ Cosmic Tracers and the Dark Energy Equation of State}

\author{M. Plionis${^{1,2}}$, R. Terlevich$^{2}$, S. Basilakos$^3$,
  F. Bresolin$^4$, E. Terlevich$^2$, J. Melnick$^5$, I. Georgantopoulos$^1$}

\address{$^1$ Institute of Astronomy \& Astrophysics, National Observatory of Athens,
    Palaia Penteli 152 36, Athens, Greece.\\
    $^2$ Instituto Nacional de Astrof\'{\i}sica Optica y Electr\'onica, AP 51
    y 216, 72000, Puebla, M\'exico.\\
    $^3$Academy of Athens, Research Center for Astronomy and Applied Mathematics,
 Soranou Efesiou 4, 11527, Athens, Greece \\
    $^4$ Institute for Astronomy of the University of Hawaii, 2680
    Woodlawn Drive, 96822 Honolulu, HI USA \\
    $^5$ European Southern Observatory, Alonso de Cordova 3107, Santiago, Chile
}

\ead{mplionis@astro.noa.gr}

\begin{abstract}
We propose to use alternative cosmic tracers to measure the dark
energy equation of 
state and the matter content of the Universe [$w(z)$ \& $\Omega_m$]. Our 
proposed method consists of two components: (a) tracing the 
Hubble relation using HII-like starburst galaxies, as an 
alternative to supernovae type Ia, which can be detected 
up to very large redshifts, $z\sim 4$, and (b) measuring the 
clustering pattern of X-ray selected AGN at a median 
redshift of $\sim 1$. Each component of the method can in itself 
provide interesting constraints on the cosmological parameters, 
especially under our anticipation that we will reduce the 
corresponding random and systematic errors significantly. 
However, by joining their likelihood functions we will be 
able to put stringent cosmological constraints and break 
the known degeneracies between the {\em dark energy} equation 
of state (whether it is constant or variable) and the matter 
content of the universe and provide a powerful and alternative 
rute to measure the contribution to the global dynamics and 
the equation of state of {\em dark energy}. 
A preliminary joint analysis of X-ray selected AGN (based on a small
XMM survey) and the currently largest SNIa sample (Kowalski et al
2008), provides: $\Omega_{\rm m}=0.28^{+0.02}_{-0.04}$ and $w=-1.0\pm 0.1$.
\end{abstract}

\section{Introduction}
We live in a very exciting period for our understanding of the
Cosmos. Over the past decade the accumulation and detailed analyses of
high quality cosmological data (eg., supernovae type Ia, CMB
temperature fluctuations, galaxy clustering, high-z clusters of
galaxies, etc.) have strongly suggested that we live in a 
flat and accelerating universe, which contains at least some sort of
cold dark matter to explain the clustering of extragalactic sources,
and an extra component which acts as having a negative pressure, as for
example the energy of the vacuum (or in a more general setting the so
called {\em dark energy}), to explain the observed accelerated cosmic
expansion  (eg. Riess, et al. 1998; 2004; 2007, Perlmutter et
al. 1999; Spergel et al. 2003, 2007, Tonry et al. 2003; Schuecker et
al. 2003; Tegmark et al. 2004; Seljak et al. 2004; Allen et al. 2004;
Basilakos \& Plionis 2005; 2006; Blake et al. 2007; Wood-Vasey et
al. 2007, Davies et al. 2007; Kowalski et al. 2008, etc).

Due to the absence of a well-motivated fundamental theory, there have
been many theoretical speculations regarding the nature of the exotic
{\em dark energy}, on whether it is a cosmological constant, a scalar or
vector fields which provide a time varying dark-energy equation of
state, usually parametrized by: 
\begin{equation}
p_Q= w(z) \rho_Q\;,
\end{equation} 
with $p_Q$ and $\rho_Q$ the pressure and density of the exotic dark
energy fluid and 
\be 
w(z)=w_0 + w_1 f(z) \;,
\label{eqstatez}
\ee 
with $w_0=w(0)$ and $f(z)$ an increasing function of redshift [ eg.,
$f(z)=z/(1+z)$] (see
Peebles \& Ratra 2003 and references therein, Chevalier \& Polarski
2001, Linder 2003, Dicus \& Repko 2004; Wang \& Mukherjee 2006). Of
course, the equation of state could be such that $w$ does not evolve
cosmologically. Current measurements do not allow us to put strong
constraints on $w$, with present limits $w \mincir -0.8$ (eg. Tonry et al. 2003;
Riess et al. 2004; Sanchez et al. 2006; Spergel et al. 2006; Wang \&
Mukherjee 2006; Davies et al. 2007).

Two very extensive recent reports have identified {\em dark energy} as a top
priority for future research: "Report of the Dark Energy Task Force 
(advising DOE, NASA and NSF) by Albrecht et al. (2006), and ``Report of
the ESA/ESO Working Group on Fundamental Cosmology'', by Peacock et
al. (2006).  It is clear that one of the most important questions in
Cosmology and cosmic structure formation is related to the nature of
{\em dark energy} (as well as whether it is the sole interpretation of the
observed accelerated expansion of the Universe) and its interpretation
within a fundamental physical theory. To this end a large number of
very expensive experiments are planned and are at various stages of
development.

Therefore, the paramount importance of the detection and
quantification of {\em dark energy} for our understanding of the cosmos and
for fundamental theories implies that the results of the different
experiments should not only be scrutinized, but alternative, even
higher-risk, methods to measure {\em dark energy} should be developed and
applied as well.

\subsection{Methods to estimate the dark energy equation of state.}
A large variety of different approaches to determine the cosmological
parameters exist. A few of the most important ones are listed below:
\begin{itemize}
\item CMB power-spectrum + Hubble relation:  By measuring the
  curvature of the universe ($\Omega_k$) from the CMB angular power spectrum
  and using SN Ia as standard candles to trace the Hubble relation
  (eg. Riess et al. 2004) to measure a combination of $\Omega_m$ and
  $\Omega_{\Lambda}$, then
  the values of all the contributing components to the global dynamics
  can be constrained (using the fact that $\Omega_k+\Omega_m+\Omega_\Lambda=1$)
\item Hubble relation + Clustering of extragalactic sources: If,
  however, the {\em dark energy} contribution is not due to a cosmological
  constant but rather it evolves cosmologically, then the Hubble
  relation provides a degenerate solution between the contribution to
  the global dynamics of the total mass and of the {\em dark energy}, even
  in the case of an almost flat geometry. In order to break this
  degeneracy one needs to introduce some other cosmological test (eg.,
  the clustering properties of galaxies, clusters or AGN, which
  can be compared with the theoretical predictions of an {\em a priori}
  selected power-spectrum of density perturbations - say the CDM - to
  constrain the cosmological parameters such as $\Omega_m$, $w$,
  $\sigma_8$ and $h$; eg., Matsubara 2004).
\item Baryonic Acoustic Oscillation method, which was identified by
  the U.S. Dark Energy Task Force as one of the four most
  promising techniques to measure the properties of the {\em dark energy}
  and the one less likely to be limited by systematic
  uncertainties. BAOs are produced by pressure (acoustic) waves in the
  photon-baryon plasma in the early universe, generated by dark matter
  (DM) overdensities. At the recombination era ($z\sim 1100$),
  photons decouple from baryons and free stream while the pressure
  wave stalls. Its frozen scale, which constitutes a “standard ruler”,
  is equal to the sound horizon length, $r_s\sim 100\; h^{-1}$ Mpc
  (e.g. Eisenstein, Hu \& Tegmark 1998). This appears as a small,
  $\sim 10\%$ excess in the galaxy, cluster or AGN power spectrum (and 2-point correlation
                                function) at the scale corresponding
                                to $r_s$. First evidences of this excess
                                were recently reported in the
                                clustering of luminous SDSS
                                red-galaxies (Eisenstein et al. 2005,
                                Padmanabhan et al. 2007). A large
                                number of photometric surveys are
                                planned in order to
                                measure {\em dark energy} (eg., the ESO/VST
                                KIDS project, DES:
                                {\tt http://www.darkenergysurvey.org},
                                Pan-STARRS:
                                {\tt http://pan-starrs.ifa.hawaii.edu}).

\item Other important cosmological tests that have been and will be
  used are based on galaxy clusters. For example, (a) the
  local cluster mass function and its evolution, $n(M,z)$, which depends
  on $\Omega_m$, on the linear growth rate of density perturbations and on the
  normalization of the power-spectrum, $\sigma_8$ (eg., Sch\"uecker et al. 2003;
  Vikhlinin et al. 2003; Newman et al. 2002; Rosati et al., 2002 and
  references therein), (b) the cluster mass-to-light ratio, $M/L$,
  which can be used to estimate $\Omega_m$ once the mean luminosity density of
  the Universe is known, assuming that mass traces light similarly
  both inside and outside clusters (see Andernach et al. 2005 for a
  recent application), (c) the baryon fraction in nearby clusters
  (eg., Fabian, 1991; White et al., 1993). Assuming that it does
  not evolve, as gasdynamical simulations indicate (eg. Gottl\"ober \&
  Yepes 2007), then its determination in distant clusters can provide
  a geometrical constraint on {\em dark energy} (Allen et al., 2004).
\end{itemize}

\subsection{Objectives of our Approach}
We wish to constrain the {\em dark energy} equation of state using the
combination of the Hubble relation and Clustering methods, but
utilizing alternative cosmic tracers for both of these components.
 
From one side we wish to trace the Hubble function using HII-like
starburst galaxies, which can be observed at higher redshifts than those
sampled by current SNIa surveys and thus at distances where the Hubble function
is more sensitive to the cosmological parameters. The HII galaxies can
be used as standard candles (Melnick, Terlevich \& Terlevich 2000,
Melnick 2003; Siegel et al. 2005) due to the correlation between their velocity
dispersion, metallicity and $H_{\beta}$ luminosity (Melnick 1978, Terlevich \&
Melnick 1981, Melnick, Terlevich \& Moles 1988). Furthermore, the use
of such an alternative high-$z$ tracer will enable us to check the SNIa based results
and lift any doubts that arise from the fact that they are the only
tracers of the Hubble relation used to-date (for possible usage of GRBs
see Ghirlanda et al. 2006; Basilakos \& Perivolaropoulos
2008)\footnote{
GRBs appear to be anything but standard candles, having a very wide
range of isotropic equivalent luminosities and energy outputs.
Nevertheless, correlations between various properties of
the prompt emission and in some cases also the afterglow emission
have been used to determine their distances.
A serious problem that hampers a straight forward use of GRBs as
Cosmological probes is the intrinsic
faintness of the nearby events, a fact which introduces a bias towards low (or high)
values of GRB observables and therefore the extrapolation
of their correlations to low-$z$ events is faced with
serious problems.
One might also expect a significant evolution of the
intrinsic properties of GRBs with redshift (also between
intermediate and high redshifts) which can be hard to disentangle
from cosmological effects. Finally, even if a reliable scaling relation
can be identified and used, the scatter in the resulting
luminosity and thus distance modulus is still fairly large.}.
We therefore plan to improve the
$L(H_{\beta})-\sigma$ distance estimator by investigating all the parameters
that can systematically affect it, like stellar age, metal and dust
content, environment etc, in order to also determine with a greater
accuracy the zero-point of the relevant distance indicator. 
The possibility to use effectively HII-like high-$z$ starburst galaxies
as cosmological standard candles, relies on our ability
to suppress significantly the present distance modulus uncertainties 
($\sigma_\mu\simeq 0.52$ mag; Melnick, Terlevich \& Terlevich 2000), which are
unacceptable large for precision cosmology.

From the other side we wish to determine the clustering pattern of
X-ray selected AGN at a median redshift of $\sim 1$, which is roughly the
peak of their redshift distribution (see Basilakos et al. 2004; 2005,
Miyaji et al. 2007). To this end we are developing the tools that will
enable us to analyse large XMM data-sets, covering up to $\sim 300$
sq.degrees of sky.


Although each of the previously discussed components of our project
(Hubble relation using HII-like starburst galaxies and angular/spatial
clustering of X-ray AGNs) will provide interesting and relatively
stringent constraints on the cosmological parameters, especially under
our anticipation that we will reduce significantly the corresponding
random and systematic errors, it is the combined likelihood of these
two type of analyses which will enable us to break the known
degeneracies between cosmological parameters and determine with great
accuracy the {\em dark energy} equation of state (see preliminary analysis
of Basilakos \& Plionis 2005; 2006).

Below we present the basic methodology of each of the two main components of
our proposal, necessary in order to constrain the {\em dark energy}
equation of state.

\section{Cosmological Parameters from the Hubble Relation}
It is well known that in the matter dominated epoch the Hubble relation depends on the cosmological
parameters via the following equation:
\be
H^2(z) = H^2_{0} \left[\Omega_m (1+z)^3 + \Omega_w \exp \left(3 \int_0^z
  \frac{1+w(x)}{1+x} {\rm d}x\right) \right]\;,
\ee
which is simply derived from Friedman's equation. We remind the reader
that $\Omega_m$ and $\Omega_w$ are the present fractional contributions to the
total cosmic mass-energy density of the matter and dark energy source
terms, respectively.

Supernovae SNIa are considered standard candles at peak luminosity
and therefore they have been used not only to determine the Hubble constant (at
relatively low redshifts) but also to trace the curvature of the
Hubble relation at high redshifts (see Riess et al. 1998, 2004, 2007; Perlmutter et
al. 1998, 1999; Tonry et al. 2003; Astier et al. 2006; Wood-Vasey et
al. 2007; Davis et al. 2007; Kowalski et al 2008). 
Practically one relates the distance modulus of the SNIa to its luminosity
distance, through which the cosmological parameters enter:
\be
\mu = m-M = 5 \log d_L + 25 \;\;\;\; {\rm where} \;\;\;\;
d_{L} =  (1+z) \int_0^z \frac{c}{H(z)} {\rm d}z \;.
\ee
The main result of numerous studies using this procedure is that distant 
SNIa's are dimmer on average by 0.2 mag than what expected 
in an Einstein-deSitter model, which translates in them being $\sim 10\%$ further
away than expected.

The amazing consequence of these results is that they imply that we live in an accelerating phase
of the expansion of the Universe, an assertion that needs to be
scrutinized on all possible levels, one of which is to verify the accelerated expansion of the
Universe using an alternative to SNIa population of
extragalactic standard candles. Furthermore, the cause and rate of the
acceleration is of paramount importance, ie., the {\em dark energy} equation of
state is the next fundamental item to search for 
 and to these directions we hope to contribute with our current project.

\subsection{Theoretical Expectations:}
To appreciate the magnitude of the Hubble relation variations 
due to the different {\em dark energy} equations of state, we plot in Figure 1 the relative
deviations of the distance modulus, $\Delta\mu$, of different {\em dark-energy}
models from a nominal {\em standard} ($w=-1$) 
$\Lambda$-cosmology (with $\Omega_m=0.27$ and
$\Omega_{\Lambda}=0.73$), with the relative deviations defined as:
\be
\Delta\mu=\mu_{\Lambda} - \mu_{\rm model} \;.
\ee
The parameters of the different models used are shown in Figure
1. As far as the {\em dark-energy} equation of state parameter is
concerned, we present the deviations from the {\em standard} model 
of two models with a constant $w$ value and of two models with an
evolving equation of state parameter, utilizing the form of eq.\ref{eqstatez}.
 In the left panel of Figure 1 we present
results for selected values of $\Omega_m$, while in the right panel
we use the same {\em dark-energy} equations of state parameters but
for the same value of $\Omega_m (=0.27)$ (ie., we eliminate the degeneracy).
\begin{figure}
\centering
\resizebox{15cm}{8cm}{\includegraphics{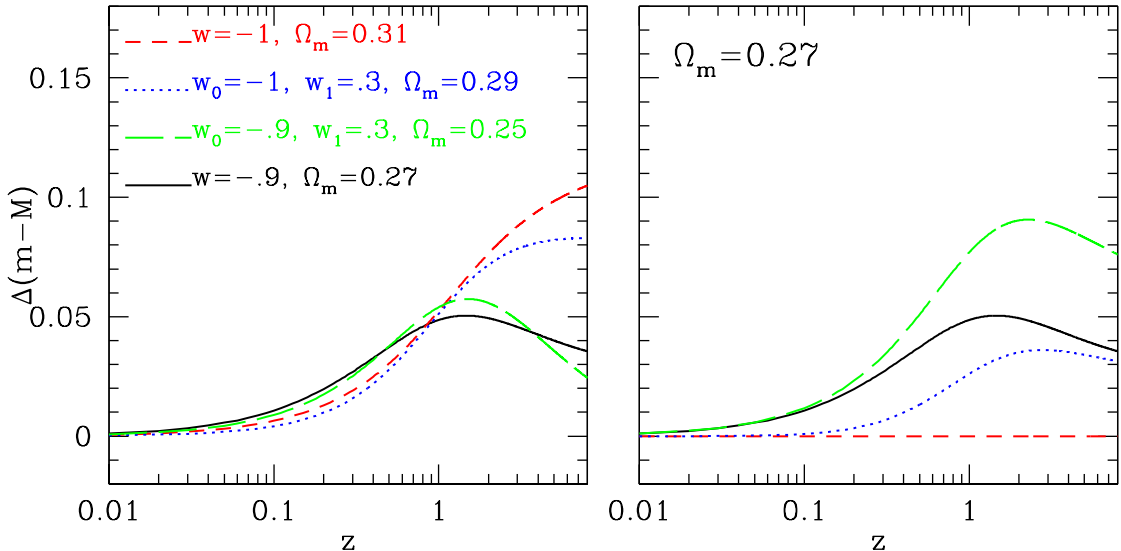}}
\caption{{\em Left Panel:} The expected distance modulus difference between the
  {\em dark-energy} models shown and the reference
  $\Lambda$-model ($w=-1$) with $\Omega_{m}=0.27$. {\em Right Panel:}
  The expected distance modulus differences once the $\Omega_m$-$w(z)$
  degeneracy is broken (imposing the same $\Omega_m$ value as the comparison model).}
\end{figure} 

Three important observations should be made from Figure 1:
\begin{enumerate}
\item The relative magnitude deviations between the different {\em dark-energy}
  models are quite small (typically $\mincir 0.1$ mag), which puts severe pressure on
  the necessary photometric accuracy of the relevant observations.
\item The largest relative deviations of the distance moduli occur at
redshifts $z\magcir 1.5$, and thus at quite larger redshifts than those
currently traced by SN Ia, and
\item There are strong degeneracies between the different cosmological
  models at redshifts $z\mincir 1$, but in some occasions even up to
  much higher redshifts (one such example is shown in Figure 1 between
  the models with $(\Omega_m,w_0,w_1)=(0.31,-1,0)$ and $(0.29,-1,0.2)$.
\end{enumerate}
Luckily, such degeneracies can be broken, as discussed
already in the introduction, by using other cosmological tests
(eg. the clustering of extragalactic sources, the CMB shift parameter, BAO's, etc). Indeed,
current evidence overwhelmingly show that the total matter content of the universe
is within the range: $0.2\mincir \Omega_m \mincir 0.3$, a fact that reduces significantly
the degeneracies between the cosmological parameters.

\subsection{Larger numbers or higher redshifts ?}
In order to define an efficient strategy to put stringent constraints
on the {\em dark-energy} equation of state, we have decided to re-analyse two recently compiled
SNIa samples, the Davies et al. (2007) [hereafter {\em D07}] compilation of 192 SNIa
(based on data from Wood-Vasey et al. 2007, Riess et al. 2007 and
Astier et al. 2007) and the
{\em UNION} compilation of 307 SNIa (Kowalski et al. 2008). Note that the two
samples are not independent since most of the {\em D07} is included in
the {\em UNION} sample.

Firstly, we present in the left panel of Figure 2 the {\em UNION} SNIa distance moduli
overploted (red-line) with the theoretical expectation of a flat cosmology
with $(\Omega_m, w)=(0.27,-1)$. In the right panel we
plot the distance moduli difference between the SNIa data and the previously
mentioned model. To appreciate the level of accuracy needed in order
to put constraints on the equation of state parameter, we also plot the
distance moduli difference between the reference $(\Omega_m,
w)=(0.27,-1)$ and the $(\Omega_m, w)=(0.27,-0.85)$ models (thin blue line).

\begin{figure}
\centering
\resizebox{16cm}{8cm}{\includegraphics{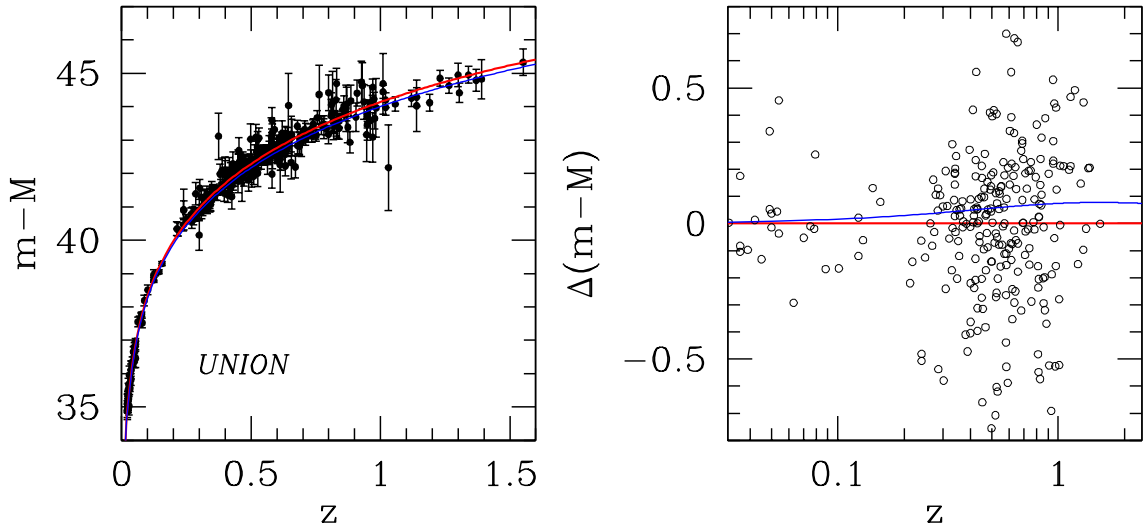}}
\caption{{\em Left Panel:} SNIa distance moduli as a function of redshift. {\em
    Right Panel:} Distance moduli difference between the
  $\Lambda$-model and the SNIa data. The blue line is the
  corresponding difference between the reference ($w=-1$) and the
  $w=-0.85$ {\em dark-energy} models.}
\end{figure}

We proceed to analyse the SNIa data by defining the usual likelihood estimator\footnote{Likelihoods
  are normalized to their maximum values.} as:
\be
{\cal L}^{\rm SNIa}({\bf c})\propto {\rm exp}[-\chi^{2}_{\rm SNIa}({\bf c})/2]
\ee
where ${\bf c}$ is a vector containing the cosmological 
parameters that we want to fit for, and
\be
\chi^{2}_{\rm SNIa}({\bf c})=\sum_{i=1}^{N} \left[ \frac{ \mu^{\rm th}(z_{i},{\bf c})-\mu^{\rm obs}(z_{i}) }
{\sigma_{i}} \right]^{2} \;\;,
\ee 
where $\mu^{\rm th}$ is given by eq.(3), $z_{i}$ is the observed redshift and $\sigma_{i}$ the observed
distance modulus uncertainty. 
Here we will constrain our analysis within the framework 
of a flat ($\Omega_{\rm tot}=1$) cosmology and therefore 
the corresponding vector ${\bf c}$ is: ${\bf c}\equiv [\Omega_m,w_0,w_1]$. We will use only SNIa with $z>0.02$ in order to avoid
possible problems related to redshift uncertainties due to the local bulk
flow.

\begin{itemize}
\item {\em Larger Numbers?} The first issue that we wish to address is how better have we done in
imposing cosmological constraints by increasing the available SNIa
sample from 181 to 292 (excluding the $z<0.02$ SNIa), ie., increasing the
sample by more than 60\%. In Table 1 we present various solutions
using each of the two previously mentioned samples. Note that since
only the relative distances of the SNIa are accurate and not their
absolute local calibration, we always marginalize with respect to the
internally derived Hubble constant (note that fitting procedures exist
which do not need to {\em a priori} marginalize over the
internally estimated Hubble constant; eg., Wei 2008).
\begin{figure}
\centering
\label{main_snia}
\resizebox{15cm}{8cm}{\includegraphics{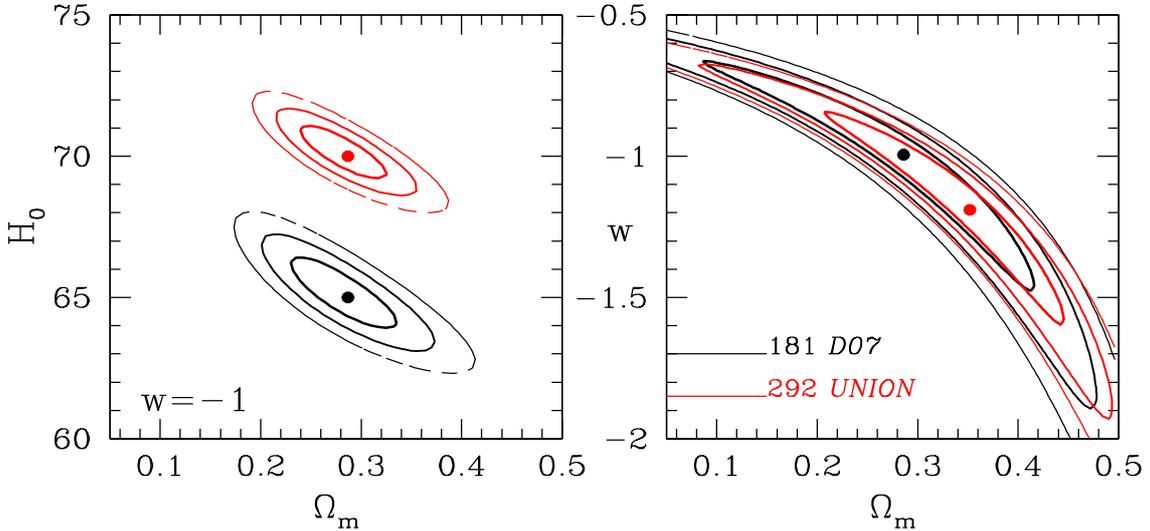}}
\caption{Solution space of the fits to the Cosmological parameters
  using either of the two SNIa data sets. The
contours are plotted where $-2{\rm ln}{\cal L}/{\cal L}_{\rm max}$ is equal
to 2.30, 6.16 and 11.83, corresponding
to 1$\sigma$, 2$\sigma$ and 3$\sigma$ confidence level.}
\end{figure} 

We present our results in Figure 3. The left panel shows the internally
derived Hubble constant for each of the to samples (and for $w=-1$).
It is evident that the derived values, which we use to free our
analysis of the $H_0$ dependence of the distance modulus, are well
constrained ($h_0=0.6515\pm 0.007$ and $0.7015\pm 0.005$ for the {\em D07} and {\em UNION}
sample, respectively). The right panel shows the main results of
interest. The size of the well-known {\em banana}
shape region of the ($\Omega_m,w$) solution space is almost identical for both samples
of SNIa. 

A first conclusion is therefore that {\em the increase by $\sim 60\%$ of the UNION sample has not
provided more stringent constraints to the cosmological
parameters.} Rather there appears to be an unexpected lateral shift of the contours
towards higher values of $\Omega_m$ and lower values of $w$, 
within, however, the 1 $\sigma$ contour of the solution space. In order to
verify that the larger number of SNIa's in the {\em UNION} sample are
not preferentially located at low-$z$'s - in which case we should have not expected more stringent
cosmological constraints using the latter SNIa sample - we plot in Figure 4 the normalized redshift frequency
distributions of the two samples (left panel) and the relative
increase of SNIa's as a function of redshift in the {\em UNION} sample with respect to the {\em D07}
sample in the right panel.
It is evident that the larger number of {\em UNION} SNIa's are
distributed in all redshifts, except for $z\magcir 1$, where there is
no appreciable increase of SNIa numbers.

\begin{figure}
\centering
\label{histz}
\resizebox{15cm}{8cm}{\includegraphics{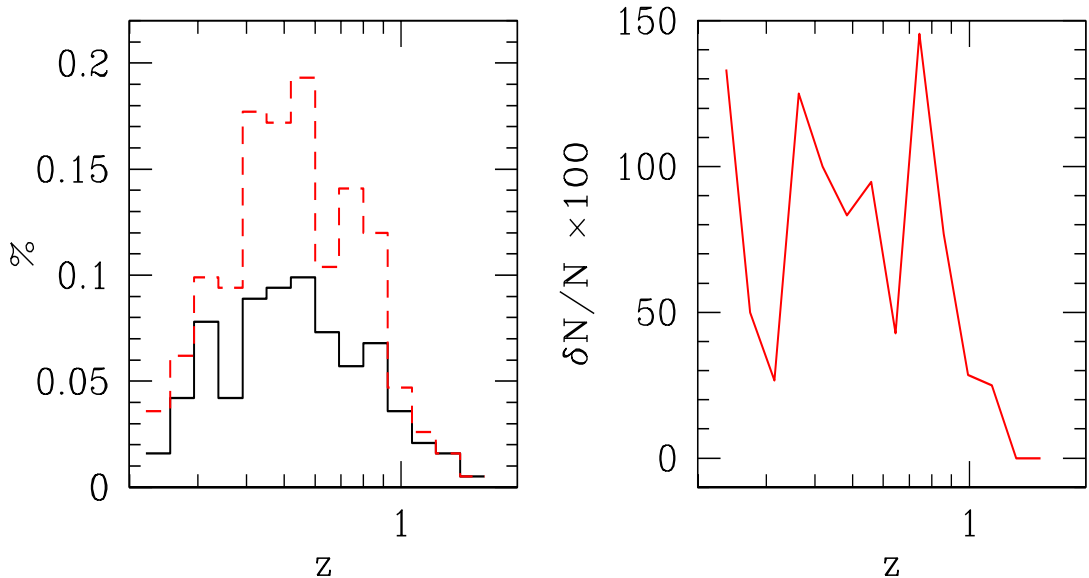}}
\caption{Redshift distribution of the {\em UNION} and {\em
    D07} SNIa's (left panel) and the relative increase of SNIa
  numbers (in percent) between the two samples.}
\end{figure}

\begin{table}
\caption{\small Fitting the SNIa data (with $z>0.02$ in order to avoid
  local bulk flow effects) to flat cosmologies. Note that for
  the case where ${\bf c}=(\Omega_m, w)$ (ie., last row), the errors
  shown are estimated after marginalizing with respect to the other
  fitted parameter.}
\tabcolsep 3pt

\begin{tabular}{|c|ccc|ccc|} \hline
      &\multicolumn{3}{c}{\em D07}  &\multicolumn{3}{c|}{\em UNION} \\ \hline 
  free params  &$w$ & $\Omega_{m}$ & $\chi^2_{\rm min}$/df & $w$ & $\Omega_{m}$ & $\chi^2_{\rm min}$/df \\ \hline
$\Omega_m$ & ${\bf -1}$ & $0.280^{+0.025}_{-0.015}$ & 187.03/180 &${\bf -1}$ & $0.280^{+0.020}_{-0.015}$ &  301.93/291 \\
$\Omega_m, w$ &$-1.025^{+0.060}_{-0.045}$ & $0.292\pm 0.018$ &
187.02/179 & $-1.212\pm 0.050$ & $0.355\pm 0.015$ & 301.11/290

 \\

\hline
\end{tabular}
\end{table}

We already have a strong hint, from the previously
presented comparison between the {\em D07} and {\em UNION}
results, that increasing the number of Hubble
relation tracers, covering the same redshift range and with the current
level of uncertainties as the available SNIa samples, 
appears to be a futile avenue in constraining further the
cosmological parameters. 

\item{\em Lower uncertainties or higher-$z$'s:}
We now resort to a Monte-Carlo procedure which will help us 
investigate which of the following two directions, which bracket many
different possibilities, would provide more stringent cosmological constraints:

\begin{itemize}
\item Reduce significantly the distance modulus uncertainties of SNIa, 
  tracing however the same redshift range as the currently available samples, or
\item use tracers of the Hubble relation located at 
  redshifts where the models show their largest relative differences
  ($z \magcir 2$), with distance modulus uncertainties comparable to that of
  the highest redshift SNIa's ($\langle \sigma_{\mu}\rangle \simeq 0.4$)
\end{itemize}

The Monte-Carlo procedure is based on using the observed high-$z$ SNIa
distance modulus uncertainty 
distribution ($\sigma_\mu$) and a model to assign random $\mu$-deviations from a
reference $H(z)$ function, that reproduces exactly the
banana-shaped contours of the $(\Omega_m,w)$ solution space of Figure
3 (right panel). Indeed, after a trial and error procedure we have
found that
by assigning to each {\em UNION} SNIa (using only their redshift) a distance
modulus deviation ($\delta \mu$) from the reference model
(in this case we use: $\Omega_m=0.35, w=-1.21$), using a Gaussian with zero mean
and variance given by observed $\langle \sigma_\mu \rangle^2$), 
and using as the relevant individual
distance modulus uncertainty, which enters as a weight in the $\chi^2$
of eq.8, the following: $\sigma^2_i=\sqrt{(1.1
\delta\mu_i)^2 + \phi^2}$, with $\phi$ a random Poisson
deviate within $[-0.01, 0.01]$, we reproduce exactly the banana-shaped
solution range of the reference model (details will appear elsewhere). This can be seen clearly in the
upper left panel of Figure 5, where we plot the original {\em UNION}
solution space (black contours) and the model solution space (red contours).
In the right panel we show the distribution of the true SNIa
deviations from the best fitted model (Table 1) as well as a random
realization of the model deviations.
\begin{figure}
\centering
\label{dev}
\resizebox{15cm}{8cm}{\includegraphics{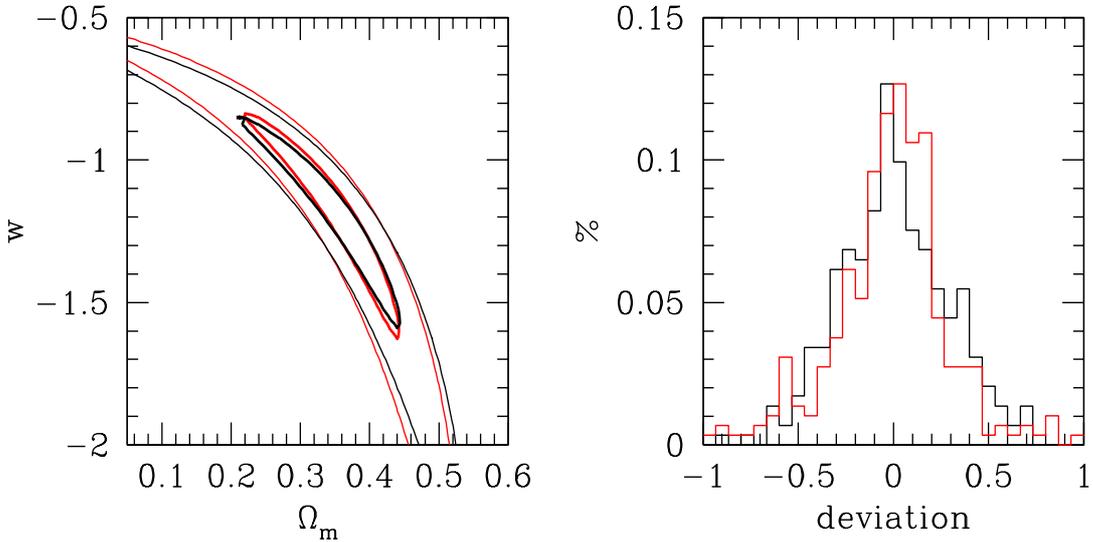}}
\caption{{\em Left Panel:} Comparison between the {\em UNION} SNIa
  constraints (black contours) and those derived by a Monte-Carlo procedure designed to closely
  reproduce them (for clarity we show only contours corresponding to 1
  and 3 $\sigma$ confidence levels). {\em Right Panel:} The {\em UNION} SNIa distance modulus
  deviations from the best fit model - $(\Omega_m,w)=(0.35,-1.21)$ -
  and a random realization of the model deviations (see text).}
\end{figure} 

Armed with the above procedure we can now address the questions posed
previously. Firstly, we reduce to half the random deviations of the
SNIa distance moduli from the
reference model (with the corresponding reduction of the relevant
uncertainty, $\sigma_i$). The results of the likelihood analysis can be seen in the
left panel of Figure 6. There is a reduction of the range of
the solution space, but indeed quite a small one. 
Secondly, we have increase artificially the high-$z$ tracers by 88
objects (by additionally using the $z>0.65$ SNIa and adding a $\delta
z=2$ to their redshift). Note that the new tracers are distributed between
$2.65\mincir z\mincir 3.55$, ie., in a range where the largest
deviations between the different cosmological models occur (see Figure 1).
The deviations from the reference model of
these additional SNIa are based on their original $\mu$ uncertainty 
 distribution (ie., we have assumed that the new high-$z$ tracers will
 have similar uncertainties as their $z\magcir 0.65$ counterparts, which
 is $\langle \sigma_\mu\rangle \simeq 0.38$).
We now find a significantly reduced solution space (right panel
of Figure 6), which shows that indeed by increasing the $H(z)$
tracers by a few tens, at those redshifts where the largest deviations
between models occur, can have a significant impact on the recovered
cosmological parameter solution space. 
\end{itemize}
\begin{figure}
\centering
\label{comp_m}
\resizebox{15cm}{8cm}{\includegraphics{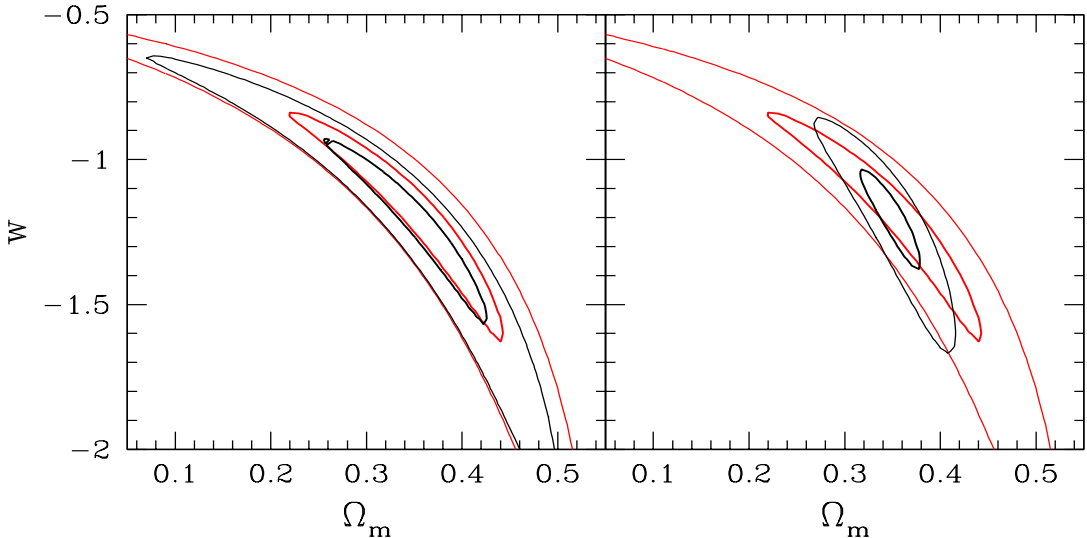}}
\caption{Comparison of the present SNIa 
  constraints (red contours) with {\em Left Panel}: those derived by reducing to half
  their uncertainties (black contours) and {\em Right Panel:}  with those derived by adding a sample of
 88 high-$z$ tracers ($2.5\mincir z \mincir 3.6$) with a
 distance modulus mean uncertainty of $\sigma_\mu\simeq 0.38$ (black contours).}
\end{figure} 

The main conclusion of the previous analysis is that the best strategy
to decrease the uncertainties of the cosmological parameters based on
the Hubble relation is to use standard candles which trace also the redshift
range $2\mincir z \mincir 4$. 
Below we present such a possibility by suggesting an alternative to the
SNIa standard candles, namely HII-like starburst galaxies (eg. Melnick
2003; Siegel et al. 2005).
 
\subsection{ Hubble Relation using HII-like starburst galaxies}
We now reach our suggestion to use an alternative and potentially 
very powerful technique to estimate cosmological
distances, which is the relation between the luminosity of the $H_{\beta}$ line and
the stellar velocity dispersion, measured from the line-widths, of HII regions
and galaxies (Terlevich \& Melnick 1981, Melnick, Terlevich \& Moles
1988). The cosmological use of this distance indicator has been tested
in Melnick, Terlevich \& Terlevich (2000) and Siegel et al (2005) (see
also the review by Melnick 2003).
It is the presence of O and B-type stars in HII regions that
causes the strong Balmer line emission, in both $H_{\alpha}$ and $H_{\beta}$. Furthermore,
the fact that the bolometric luminosities of HII galaxies are
dominated by the starburst component implies that their luminosity per
unit mass is very large, despite the fact that the galaxies are
low-mass. Therefore they can be observed at very large redshifts, and
this fact makes them cosmologically very interesting
objects. Furthermore, it has been shown that the $L(H_{\beta})-\sigma$ correlations
holds at large redshifts (Koo et al. 1996, Pettini et al. 2001, Erb et
al. 2003) and therefore they can be used to trace the Hubble relation
at cosmologically interesting distances. One of the most important
prerequisites in using such relations, as distance estimators, is the
accurate determination of their zero-point. To this end, Melnick et al
(1988) used giant HII regions in nearby late-type galaxies and derived
the following empirical relation (using a Hubble constant of $H_0=71$
km/sec/Mpc):

\begin{equation}
\log L(H_{\beta}) = \log M_z + 29.60 \;\; {\rm with} \;\;  M_z=\sigma^5/(O/H)
\end{equation}
where $O/H$ is the metallicity. Based on the above relation and the work
of Melnick, Terlevich \& Terlevich (2000), the distance modulus of HII
galaxies can be derived from:
\begin{equation}
\mu = 2.5 \log(\sigma^5/F_{H\beta})- 2.5 \log(O/H)- A_{H\beta} -26.44
\end{equation}
with $F_{H\beta}$ and $A_{H\beta}$ are the flux and extinction in
$H_{\beta}$. The rms dispersion
in distance modulus was found to be $\sim 0.52$ mag.
The analysis of Melnick, Terlevich \& Terlevich (2000) has
shown that most of this dispersion ($\sim 0.3$ mags)
comes from observational errors in the stellar velocity dispersion measurements, 
from photometric errors and metallicity effects. 
It is therefore important to understand and correct the sources of
random and systematic errors of the $L(H_{\beta})-\sigma$ relation, and
indeed with the availability of new observing
techniques and instruments, we hope to reduce significantly the
previously quoted rms scatter.

A few words are also due to the possible systematic effects of the
above relation. Such effects may be related to the age of the HII
galaxy (this can be dealt with by putting a limit in the equivalent
width of the $H_{\beta}$ line, eg. $EW(H_{\beta})>$ 25 Angs; see
Melnick 2003), to extinction, to different
metallicities and environments. Also the $EW(H_{\beta})$ of HII-like galaxies at intermediate and
high redshifts are smaller than the local ones, a fact which should be
taken into account.

We have commenced an investigation of all these
effects by using high-resolution spectroscopy of a relatively large
number of SDSS low-$z$ HII galaxies, with a
range of $H_{\beta}$ equivalent widths, luminosities, metal content and local
overdensity,  in order
to reduce the scatter of the HII-galaxy based distance estimator to
about half its present value, ie., our target is $\sim 0.25$ mag.
We will also define a medium and
high redshift sample (see Pettini et al. 2001; Erb et al. 2003),
consisting of a few hundred objects distributed in the different
high-redshift bins, which will finally be used to define the high-$z$
Hubble function. 
\begin{figure}
\label{cosmoHII}
\centering
\resizebox{8cm}{8cm}{\includegraphics{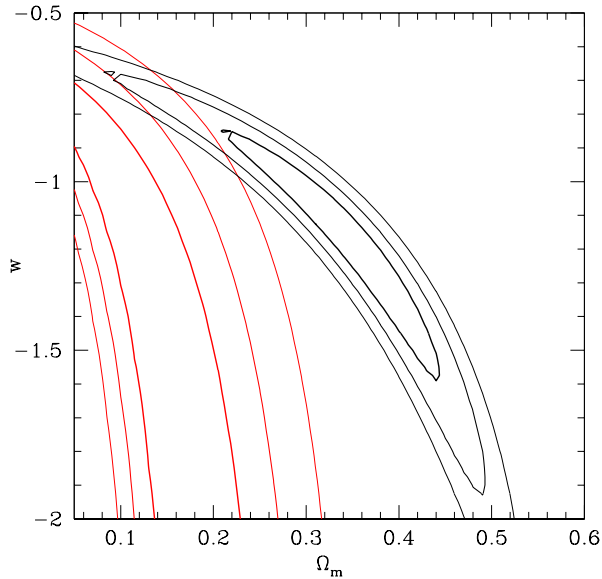}}
\caption{Comparison between the {\em UNION} SNIa
  constraints and those derived by using the 15 high-$z$ starburst
  galaxies of Siegel et al. (2005). Although, the latter constraints
  are weak and mostly inconsistent with the former, this plot serves to
  indicate the potential of using high-$z$ starburst galaxies (once of
  course we have reduced significantly their distance modulus uncertainties).}
\end{figure} 

{\em Summarizing, the use of HII galaxies to trace the Hubble relation, as an
alternative to the traditionally used SN Ia, is based on the following
facts: 
\begin{itemize}
\item[(a)] local and high-$z$ HII-like galaxies and HII regions are
physically very similar systems (Melnick et al 1987) providing a
phenomenological relation between the luminosity of the $H_{\beta}$ line, the
velocity dispersion and their metallicity as traced by $O/H$ (Melnick,
Terlevich \& Moles 1988). Therefore HII-like starburst galaxies can be
used as alternative standard candles (Melnick, Terlevich \& Terlevich
2000, Melnick 2003; Siegel et al. 2005) 
\item[(b)] such galaxies can be readily observed at
much larger redshifts than those currently probed by SNIa and 
\item[(c)] it is at such higher redshifts that the differences between the
predictions of the different cosmological models appear more vividly.
\end{itemize}
}

Already a sample of 15 such high-z
starburst galaxies have been used by Siegel et al. (2005) in an
attempt to constrain
cosmological parameters but the constraints, although in the correct
direction, are very weak. 
Here we perform our own re-analysis of this data-set and the resulting
constraints on the $\Omega_m,w$ plane (for a flat geometry) can be
seen in Figure 7.
Note that imposing $w=-1$, our analysis of the Siegel et al (2005) data set
 provides $\Omega_m=0.10\pm0.05$, which is towards the lower side of
 the generally accepted values.
Comparing these HII-based results to the present
constraints of the latest SNIa data ({\em D07} and {\em UNION})
clearly indicates the necessity to:
\begin{itemize}
\item re-estimate carefully the local zero-point of the
  $L(H_{\beta})-\sigma$ relation,
\item suppress the HII-galaxy distance modulus uncertainties, 
\item increase the high-$z$ starburst sample by a large fraction,
\item make sure to select high-$z$ bona-fide HII-galaxies, excluding those that
  show indications of rotation (Melnick 2003). 
\end{itemize}

\section{The Clustering of high-z X-ray AGN}
X-ray selected AGNs provide a relatively unbiased census of the
AGN phenomenon, since obscured AGNs, largely missed in optical surveys,
are included in such surveys.
Furthermore, they can be detected out to high redshifts and thus trace
the distant density fluctuations providing important
constraints on super-massive black hole formation,
the relation between AGN activity and Dark Matter (DM) halo hosts,
the cosmic evolution of the AGN phenomenon (eg. Mo \& White 1996,
Sheth et al. 2001), and on cosmological parameters and the dark-energy
equation of state (eg. Basilakos \& Plionis 2005; 2006).

The earlier ROSAT-based analyses
(eg. Boyle \& Mo 1993; Vikhlinin \& Forman 1995; Carrera et al. 1998;
Akylas, Georgantopoulos, Plionis, 2000; Mullis et al. 2004)
provided conflicting results on the nature and amplitude of high-$z$ AGN
clustering.
With the advent of the XMM and {\em Chandra} X-ray observatories,
many groups have attempted to settle this issue, but in vain.
Different surveys have provided
again a multitude of conflicting results, intensifying the debate
(eg. Yang et al. 2003; Manners et al. 2003;
Basilakos et al. 2004; Gilli et al. 2005; Basilakos et al 2005;
Yang et al. 2006; Puccetti et al. 2006; Miyaji et al. 2007; Gandhi et al.
2006; Carrera et al. 2007). However, the recent indications of a flux-limit dependent
clustering appears to remove most of the above inconsistencies (Plionis et al. 2008;
see also Figure 8).

\begin{figure*}
\centering
\resizebox{14cm}{8cm}{\includegraphics{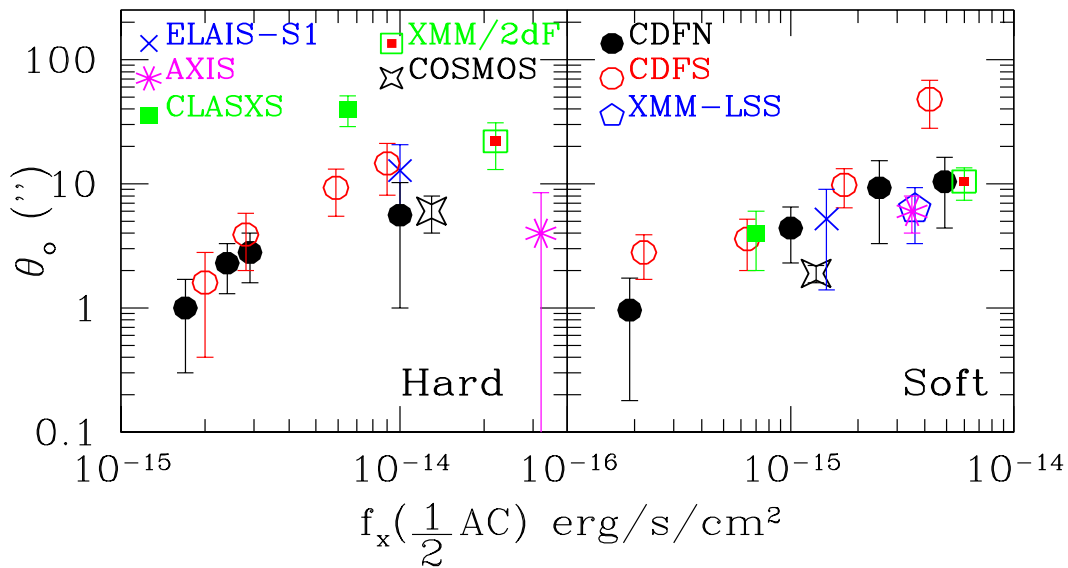}}
\caption{The angular correlation scale, $\theta_0$, as a function of
different survey characteristic flux, defined as that corresponding to half the
respective survey area-curves. Most results appear to be consistent with
the clustering flux-limit dependence, found from the CDF-N and
CDF-S (from Plionis et al. 2008).}
\end{figure*} 

Furthermore, there  are indications for a quite large high-$z$ AGN clustering length,
reaching values $\sim 15 - 18 \;h^{-1}$ Mpc at the brightest flux-limits (eg.,
Basilakos et al 2004; 2005, Puccetti et al. 2006, Plionis et al. 2008), a fact which, if verified,
has important consequences for the AGN bias evolution and therefore for the evolution of the
AGN phenomenon (eg. Miyaji et al. 2007; Basilakos, Plionis \& Ragone-Figueroa 2008).
An independent test of these results will be to establish that the environment of high-$z$
AGN is associated with large DM haloes, which being massive should be more clustered.

Below we justify the necessity for a large-area XMM survey
in order to unambiguously determine
the clustering pattern of high-$z$ ($z\sim 1$) X-ray AGNs.
 We further show that such measurements
can be used to put strong cosmological constraints (see for example
Basilakos \& Plionis 2005; 2006), and help break the $\Omega_m, w$ degeneracies.

\subsection{Biases and Systematics:}
It is important to understand and overcome the shortcomings and problems that one is facing in
order to reliably and unambiguously determine the clustering properties of the X-ray
selected AGNs. Below we list the source of such problems, some of which can be rectified
by considerably increasing the X-ray survey area.
\begin{itemize}
\item {\em Cosmic Variance}: Is the volume surveyed large enough to smooth out inhomogeneities
of the large-scale distribution of AGNs? (for example see Stewart et al. 2007).
Closely related to this problem is the so-called integral
constraint, which practically depends on the unknown true mean density of the cosmic sources under study.
If the area is small enough, then the mean density, estimated from the survey itself,
is way-out of its true value and thus the
usual correlation function analysis will impose the observed mean number density as the true one
(an example of this is the CDF-S were a large number of superclusters at $z\sim 0.7$ are found;
see Gilli et al. 2003).
This usually results into an underestimation of the true correlation amplitude and a shallower
zero-crossing of the estimated $\xi(r)$ or $w(\theta)$.  
A source of the observed scatter between the presently available surveys (see Figure 8) could well
be the {\em cosmic variance}. These problems, however, are rectified with the large-area
XMM survey proposed. 
\item {\em The amplification bias} which can enhance
artificially the clustering signal due to the detector's PSF smoothing
of source pairs with intrinsically small angular separations
(see Vikhlinin \& Forman 1995; Basilakos et al. 2005).
This problem can affect clustering results if at the median
redshift of the sources under study the XMM PSF angular size
corresponds to a rest-frame
spatial scale comparable to the typical source pair-wise separations.
Furthermore, the possible variability of the PSF size through-out the XMM fields
can have an additional effect. This should be modeled and tested with Monte-Carlo simulations
in order to establish the extent to which the clustering results are affected. In large-area
surveys it is necessary to take good-care of this effect when using
source pairs that belong to different XMM pointings.

\item {\em Reliable production of random source catalogues}:
This is an issue which is extremely important and not appreciated at the necessary extent.
The random catalogues, with which the observed source-pairs are compared, should be
produced to account for all systematic effects from which the observations suffer, among which
the different positional sensitivity and edge effects of each XMM pointing.
Furthermore, a reliable $\log N - \log S$ distribution
(theoretically motivated or observationally determined) should be reproduced in the random
``XMM pointings''
and the random sources should be observed following the same procedure as in the true observations.
Random positioned sources with fluxes lower
than that corresponding to the particular position of the sensitivity
map of the particular XMM pointing should be removed from each random catalogue.
\end{itemize}

An optimal approach to unambiguously determine the clustering pattern of X-ray selected AGNs
would be to determine both the angular and spatial clustering pattern.
The reason being that various systematic effects or uncertainties enter differently in the
two types of analyses. On the one side, using $w(\theta)$ and its Limber inversion, one by-passes
the effects of redshift-space distortions and uncertainties related to possible misidentification
of the optical counter-parts of X-ray sources. On the other side,
using spectroscopic or accurate photometric redshifts
 to measure $\xi(r)$ or $w_p(\theta)$ one by-passes
the inherent necessity, in Limber's inversion of $w(\theta)$, of the source
redshift-selection function (for the determination of which one uses the
integrated X-ray source
luminosity function, different models of which exist). For the inversion to work it is also
necessary to model the spatial correlation function as a power
law, to assume a clustering
evolution model, which is taken usually to be that of constant clustering in comoving
coordinates (eg. de Zotti et al. 1990; Kundi\'c 1997) and also to assume a cosmological model.
 Limber's inversion then reads:
\begin{equation}
\theta_{\circ}^{\gamma-1}=H_{\gamma}r_{\circ}^{\gamma}
\left(\frac{H_\circ}{c}\right)^\gamma
\int_{0}^{\infty} \left( \frac{1}{N}\frac{{\rm d}N}{{\rm d}z}
\right)^{2} \frac{E(z)}{x^{\gamma-1}(z)}  
{\rm d}z \;,
\end{equation}
where $\epsilon=\gamma-3$ for the constant clustering in comoving coordinates model,
$x(z)$ is the proper distance,
$E(z)=\sqrt{\Omega_{\rm m}(1+z)^{3}+\Omega_{\Lambda}}$ and
$H_{\gamma}=\Gamma(\frac{1}{2}) \Gamma(\frac{\gamma-1}{2})/\Gamma
(\frac{\gamma}{2})$.
As noted previously, for the inversion to be possible it is necessary to know
the X-ray source redshift distribution, ${\rm d}N/{\rm d}z$, and the total number, $N$, of the X-ray
sources, which can be determined by integrating the
X-ray source luminosity function above the minimum
luminosity that corresponds to the particular flux-limit used.

\subsection{X-ray surveys}
In order to reach a flux-limit for which the soft-band clustering appears
to converge to its final value (due to the flux-limit-clustering correlation, revealed in
Plionis et al. 2008; see also Figure 8) we plan to analyse all $>$10 ksec XMM pointings, which
will allow us to reach a flux-limit of $\sim 2 \times 10^{-15}$ erg/sec/cm$^{2}$ in the
soft (0.5-2 keV) and $\sim 10^{-14}$ erg/sec/cm$^{2}$ in the hard (2-10 keV) bands, respectively.
Such an exposure time will finally provide (taking in to account
realistci observational effects) $\sim$250 soft-band and $\sim$100 hard-band X-ray sources per deg$^2$,
according to the Kim et al. (2007) $\log N-\log S$ and since
all publicly available XMM pointings, with exposure time more than 10 ksec,
add to $\sim 300$ non-contiguous sq.degrees, implies a resulting
sample of $\sim 75000$ soft and $\sim 30000$ hard X-ray sources. These
numbers will allow us to derive
with great accuracy the small-separation angular correlation function,
the Limber's inversion of which can provide their spatial correlation
function.
Furthermore, around 100 non-contiguous sq.degrees of the previous survey (except for
a few contiguous regions with areas between 2 and 10 sq.degrees each)
are covered also by the SDSS, providing crude photo-$z$’s. This will
allow us to derive the angular correlation function in distinct redshift
bins in the range $0.5<z<2$ and thus quantify the evolution of the bias
of X-ray selected AGN, an important ingredient in disentangling the
cosmological parameters. Note finally that $\sim$50 sq.degrees are 
covered also by the publicly available UKIDSS ({\tt http://www.ukidss.org/})
which provide 3 near-IR colours and thus for a subsample of the
previous data we will have relatively more accurate photo-$z$’s
allowing us to attempt to derive directly the spatial correlation
function.

{\em Summarizing, the analysis of such large XXM surveys will allow us
to unambiguously determine the soft and hard-band X-ray AGN
clustering pattern, minimizing the biases and systematic effects 
discussed in the previous section, as well as to 
study the evolution of the AGN correlation function (utlizing photo-$z$'s
and dividing the angular sample into 2-3 $z$-bins)}.



Finally, together with a large European consortium, we are planning to
survey a 2-3 contiguous sky areas adding to $\sim 50$ deg$^2$ and
covered also by
a large number of other multiwavelength surveys (like UKIDSS, NEWFIRM, etc),
which will probably
allow us to sample the long wavelength regime of the AGN correlation function, and
thus measure Baryonic Accoustic Oscillations, which are extremely
important for Dark-energy investigations (Peacock et al 2006; Albrecht et al. 2006).
To obtain the necessary high photo-$z$ accuracy, one may envision a wide-field
multi-filter (say using $\sim$20 broad and narrow-band filters)
survey, specially tuned in order to obtain
relatively high-accuracy photo-$z$'s of $z\sim 1$ AGN. Thoughts for
the aquisition of such an instrument, to
be mounted on the 2.3m Hellenic Aristarchos telescope, are already
been discussed in the Institute of Astronomy \& Astrophysics of the
National Observatory of Athens.

\subsection{Cosmological Parameter constraints from X-ray AGN Clustering}

\begin{figure}
\label{bias}
\center{
\resizebox{14cm}{8cm}{\includegraphics{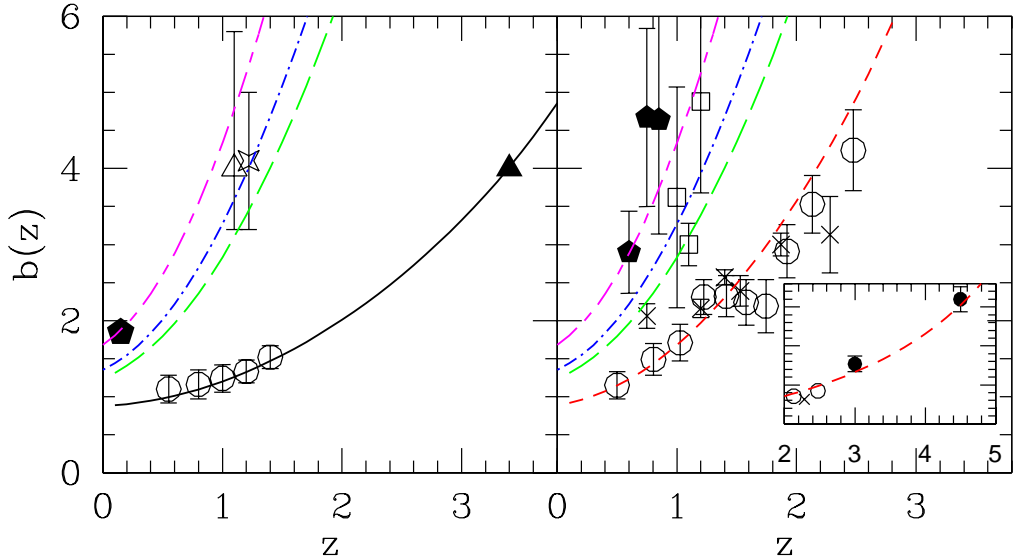}}}
\caption{
Comparison of the Basilakos et al (2008) $b(z)$ evolution
model with different observational data. Different line types represent different
halo masses.
{\em Left Panel:} optical galaxies (open points) with solid line
corresponding to $M_{\rm DM}\sim 10^{12} \; h^{-1} \; M_{\odot}$,
Lyman break galaxies (solid triangle), EROs (star), DRGs (open triangle)
and 2dF radio sources (filled pentagon). The dot-dashed line corresponds
to 7.7 $\times 10^{13} \;h^{-1} \; M_{\odot}$.
{\em Right Panel:} optically selected quasars (open points and crosses),
soft and hard X-ray point sources (open squares and solid
diamonds; the large scatter corresponds to the uncertainty of the present day
clustering results; see Plionis et al. 2008 and references therein).
In the insert we plot, as solid points, the high-$z$ SSRS
DR5 QSOs and the same $b(z)$
model that fits their lower redshift counterparts (ie., $M_{\rm DM} \simeq
10^{13} \;h^{-1} \; M_{\odot}$).
}
\end{figure}

The unambiguous determination of the correlation function of the $z\sim1$
X-ray AGNs, even in angular space, will allow us to estimate with
good precision (a) their relation to the underlying matter
fluctuations at $z\sim 1$ (ie., their bias), (b)
the evolution of their bias and therefore the mass of the DM
haloes which they inhabit (eg., Miyaji et al. 2007; Basilakos,
Plionis \& Ragone-Figueroa 2008) and (c) put strong cosmological
constraints on the $\Omega_m, h$ or $\Omega_m, \sigma_8$ planes,
while with the help of the high-$z$ HII-based Hubble relation 
on the $\Omega_m, w(z)$ space.

It is well known  (Kaiser 1984; Benson et al. 2000) that according to
linear biasing the correlation function of the AGN (or any mass-tracer)
($\xi_{\rm AGN}$) and dark-matter one ($\xi_{\rm DM}$), are related by:
\be
\label{eq:spat}
\xi_{\rm AGN}(r,z)=b^{2}(z) \xi_{\rm DM}(r,z) \;\;,
\ee
where $b(z)$ is the bias evolution function (eg.
Mo \& White 1996, Matarrese et al. 1997, Basilakos \& Plionis
2001; 2003; Basilakos, Plionis \& Ragone-Figueroa 2008)

A first outcome of the proposed correlation function analysis will be the
accurate determination
of the AGN bias at their median redshift (in our case $\bar{z} \simeq 1$), utilizing:
$$
b(z)=\left(\frac{ r_{0} }{r_{0,m}} \right)^{\gamma/2}
D^{3+\epsilon}(z)\;\;\; \mbox{with} \; \gamma=1.8 \; {\rm and} \;\; \epsilon=-1.2 \;,
$$
where $r_0$ and $r_{0,m}$ are the measured AGN and dark-matter (from
the $P(k)$)
clustering lengths, respectively, while $D(z)$ is the perturbation's linear growing
mode.
Then using a bias evolution model, one will be able to determine the mass of the DM halo within which
such AGN live (eg. see Figure 9). 
\begin{figure}
\label{X-anal}
\resizebox{15cm}{7.5cm}{\includegraphics{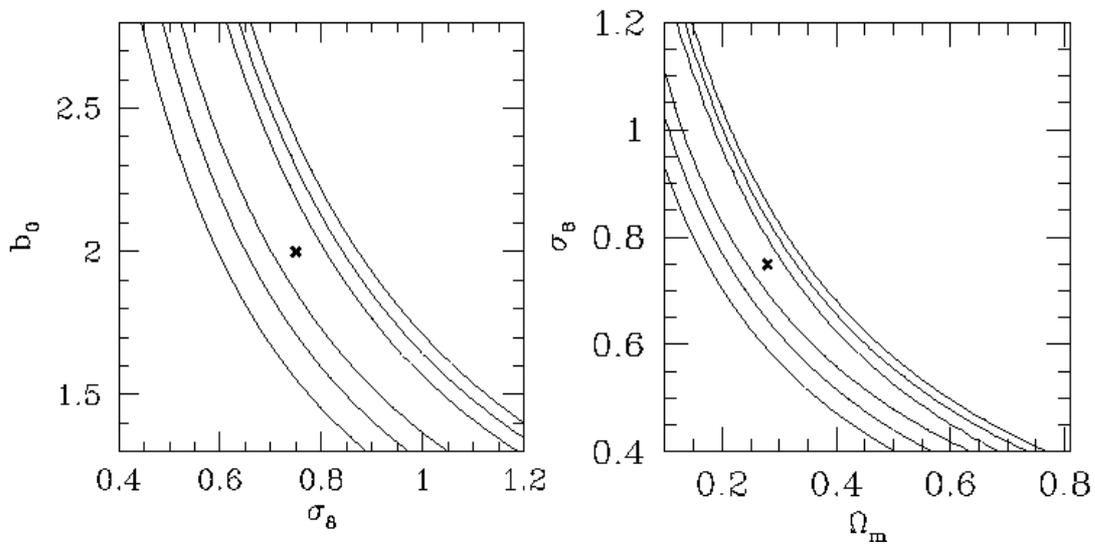}}
\caption{Likelihood contours from the X-ray AGN clustering analysis of
  Basilakos \& Plionis (2005) in the following planes:
$(\Omega_m,\sigma_{8})$ (right panel),
$(\sigma_{8},b_{0})$ (left panel). The
contours plotted correspond
to 1$\sigma$, 2$\sigma$ and 3$\sigma$ confidence level.}
\end{figure}

Furthermore, we will compare the observed AGN clustering
with the predicted, for different cosmological models,
correlation function of the underlying mass,
$\xi_{\rm DM}(r,z)$. To this end we can use
the Fourier transform of the spatial power spectrum $P(k)$:
\be
\label{eq:spat1}
\xi_{\rm DM}(r,z)=\frac{(1+z)^{-(3+\epsilon)}}{2\pi^{2}}
\int_{0}^{\infty} k^{2}P(k)
\frac{{\rm sin}(kr)}{kr}{\rm d}k \;\;,
\ee
where $k$ is the comoving wavenumber, $P(k) =P_{0} k^{n}T^{2}(k)$ the CDM power-spectrum
with scale-invariant ($n=1$) primeval inflationary fluctuations, while
the transfer function parameterization is as in
Bardeen et al. (1986), with the corrections given approximately
by Sugiyama (1995).

Basilakos \& Plionis (2005; 2006) have already used succesfully
a standard maximum likelihood procedure to compare the measured
XMM source angular correlation function from a relatively small ($\sim
2$ sq.degrees survey; Basilakos et al. 2005)
with the prediction of different spatially flat cosmological models,
and derived interesting cosmological contraints (for flat and
constant-$w$ cosmologies).

In Figure 10 we present the constraints provided from the
 preliminary analysis by Basilakos \& Plionis (2006) of a $\sim 2$
 sq.degrees XMM survey, on the present bias-factor ($b_0$) of the X-ray AGN, on the
normalization of the power spectrum ($\sigma_8$) and on $\Omega_m$
(marginalizing over different parameters).
In Figure 11 (left panel) we present the $(\Omega_m, w)$ constraints provided by the X-ray
AGN clustering analysis (red contours), once we have marginalized over 
the $\sigma_{8} (\sim 0.8)$ and the bias factor at the present time
($\sim 2$).

\section{Joint Hubble-relation and Clustering analysis}
It is evident from Figure 11 (left panel) that $w$ is degenerate with respect to $\Omega_m$
and that all the values in the interval $-2\le w \le -0.35$ are acceptable
within the $1\sigma$ uncertainty.
However, we break this degeneracy by adding the constraints provided
by the Hubble relation technique, using here the {\em UNION} SNIa
sample.
We therefore perform a joint likelihood analysis, assuming that the two data
sets are independent (which indeed they are) and thus we can write the
joint likelihood as the product of the two individual ones.
\begin{figure}
\label{X-SN}
\centering
\resizebox{15cm}{9cm}{\includegraphics{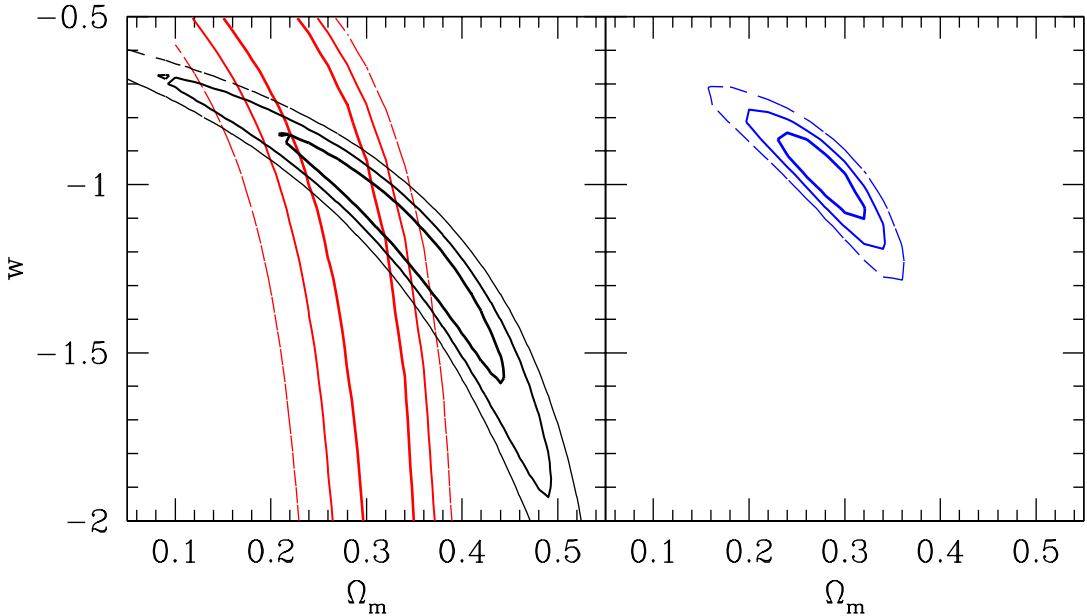}}
\caption{{\em Left Panel:} Likelihood contours on the $\Omega_m, w$ plane 
from the X-ray AGN clustering analysis of Basilakos \& Plionis (2005;
2006) (red contours) and the {\em
  UNION} SNIa analysis (black contours). {\em Right Panel:} The Joint
likelihood contours.}
\end{figure}

Our current joint likelihood analysis, once we impose $h=0.72$ and
$\sigma_8=0.8$, provides quite stringent constraints of the $\Omega_m$
and $w$ parameters:
$$\Omega_{\rm m}=0.28^{+0.02}_{-0.04} \;\;\;\; {\rm and} \;\;\;\; w=-1.0\pm 0.1\;.$$
However, the uncertainty on $w$ is still quite large, while the
necessity to impose constraints on a more general, time-evolving, {\em dark-energy} equation 
of state (eq. 2) implies that there is ample space for great
improvment and indeed the aim of the project, detailed in these
proceedings, is exactly in this direction.



\end{document}